\documentclass[draft]{agujournal2019}
\usepackage{url} 
\usepackage{lineno}
\usepackage{subcaption}
\usepackage{graphicx}
\usepackage{trackchanges}

\usepackage{float}
\usepackage{amsfonts} 

\draftfalse

\journalname{AGU Journals}

\begin{document}

\title{Automated Classification of Plasma Regions at Mars Using Machine Learning}

\authors{
Yilan Qin\affil{1,2}, 
Chuanfei Dong\affil{1,2,3}, 
Hongyang Zhou\affil{2}, 
Chi Zhang\affil{2},
Kaichun Xu\affil{1}, 
Jiawei Gao\affil{2}, 
Simin Shekarpaz\affil{2,4},
Xinmin Li\affil{2}, 
Liang Wang\affil{2}
}

\affiliation{1}{Department of Electrical and Computer Engineering, Boston University, Boston, MA 02215, USA}
\affiliation{2}{Center for Space Physics, Boston University, Boston, MA 02215, USA}
\affiliation{3}{School of Natural Sciences, Institute for Advanced Study, Princeton, NJ 08540, USA}
\affiliation{4}{Department of Mathematics and Statistics, Boston University, Boston, MA 02215, USA}

\begin{keypoints}
\item Machine learning models using MAVEN ion energy spectra are developed to automatically classify plasma regions at Mars.
\item A CNN exploiting short-timescale spectral continuity achieves $\sim$95\% accuracy and outperforms an MLP.
\item The CNN approach enables efficient identification of plasma regions under varying solar wind conditions for MAVEN and Mars missions.

\end{keypoints}

\begin{abstract}

The plasma environment around Mars is highly variable because it is strongly influenced by the solar wind. Accurate identification of plasma regions around Mars is important for the community studying solar wind–Mars interactions, region-specific plasma processes, and atmospheric escape. In this study, we develop a machine-learning–based classifier to automatically identify three key plasma regions—solar wind, magnetosheath, and induced magnetosphere—using only ion omnidirectional energy spectra measured by the MAVEN Solar Wind Ion Analyzer (SWIA). Two neural network architectures are evaluated: a multilayer perceptron (MLP) and a convolutional neural network (CNN) that incorporates short temporal sequences. Our results show that the CNN can reliably distinguish the three plasma regions, whereas the MLP struggles to separate the solar wind and magnetosheath. Therefore, the CNN-based approach provides an efficient and accurate framework for large-scale plasma region identification at Mars and can be readily applied to future planetary missions.
\end{abstract}

\section*{Plain Language Summary}

Mars does not have a global intrinsic magnetic field like Earth. As a result, its upper atmosphere interacts directly with the solar wind—a stream of fast-moving plasma continuously emitted from the Sun—along with the magnetic field carried within it. This interaction forms a global induced magnetosphere around Mars. Because the solar wind is highly variable, the electromagnetic fields and plasma environment around Mars are also highly variable and typically consist of several distinct plasma regions. Identifying these regions is important for scientists studying how the solar wind interacts with Mars and the associated magnetospheric processes. To help address this problem, we develop a machine-learning tool that can automatically identify plasma regions using observations from the MAVEN spacecraft. By comparing with ground-truth results, we show that the developed machine learning tool is accurate and robust, and it can also be readily applied to future planetary missions such as ESCAPADE.

\section{Introduction}

Unlike Earth, Mars lacks a global intrinsic dipole magnetic field, and therefore the solar wind can directly interact with the upper atmosphere and ionosphere. This direct interaction allows the solar wind to efficiently remove atmospheric ions and is considered an important driver of atmospheric escape at Mars \cite{Dong2014,Dong2015,Dong2018a,Dong2018b,Jakosky2018,Lillis2015,Ramstad2018,Zhang2025a}. The solar wind interaction with Mars generates an induced magnetosphere through the draping and pileup of the interplanetary magnetic field (IMF) around the planet \cite{Luhmann2004,Halekas2021,Ramstad2020,Zhang2022}. This induced magnetosphere includes major plasma boundaries and regions such as the bow shock, magnetosheath, and induced magnetosphere, forming a space environment that in some respects resembles the intrinsic magnetospheres of magnetized planets \cite{Nagy2004}. However, unlike intrinsic magnetospheres, the Martian induced magnetosphere is highly sensitive to upstream conditions, such as the IMF and solar wind dynamic pressure \cite{Cheng2025}. Variations in these external drivers can significantly modify the electromagnetic environment around Mars, thereby influencing planetary ion acceleration, transport, and escape \cite{Dubinin2017,Inui2018,Zhang2024}, as well as electron dynamics \cite{Zhang2025b}. Therefore, studies of solar wind interaction with Mars, atmospheric escape, and related plasma processes must explicitly consider upstream solar wind conditions.

However, due to the limited spatial and temporal coverage of spacecraft observations, continuous measurements of the upstream solar wind at Mars are generally unavailable. Consequently, researchers must manually identify time intervals when the spacecraft is located in the upstream region in order to determine the background solar wind conditions. This manual procedure significantly reduces the efficiency of data analysis. To address this limitation, several automatic methods have been developed to distinguish plasma regions around Mars. \citeA{Halekas2017} proposed an approach based on ion bulk parameters and magnetic field measurements. \citeA{Nemec2020} developed an empirical model using ion bulk parameters, which requires upstream solar wind parameters for normalization. Building on the work of \citeA{Nemec2020}, \citeA{Linzmayer2024} developed a machine-learning approach for automatic region identification. However, these methods generally rely on multiple input parameters, such as magnetic field measurements and ion bulk properties, and involve relatively complex processing procedures.

Recent advances in machine learning have provided a powerful new approach for automatic plasma-region classification. In intrinsic magnetospheres, convolutional neural networks have been used to identify plasma regions in near-Earth space using MMS 3D particle energy distributions~\cite{Olshevsky2021}, and random forest classifiers have been applied to classify magnetospheric regions and identify boundary crossings at Mercury using MESSENGER magnetometer and ephemeris data~\cite{Hollman2026}. At Saturn, convolutional neural networks have been applied to detect major plasma boundaries, such as the bow shock and magnetopause, using Cassini magnetic field data and electron energy--time spectrograms~\cite{Cheng2022}. These studies demonstrate that plasma regions and boundaries can be distinguished using relatively simple inputs, such as particle energy distributions, magnetic field measurements, and spacecraft position information, without requiring derived plasma parameters. This capability makes machine learning approaches particularly suitable for plasma-region classification.

Motivated by these developments, we apply a machine learning approach to plasma-region classification at Mars. The Mars Atmosphere and Volatile EvolutioN (MAVEN) mission \cite{Jakosky2015} has provided more than a decade of observations spanning both upstream and downstream regions, together with high time-resolution plasma measurements, forming a comprehensive dataset for machine learning applications. Based on this dataset, we develop an automatic plasma-region classifier using a convolutional neural network (CNN) that incorporates temporal variations by using consecutive ion omnidirectional energy spectra. Unlike previous methods at Mars that rely on multiple parameters, the classifier uses only ion energy spectra as input. The model achieves an accuracy of approximately 95\% and outperforms a multilayer perceptron (MLP) baseline. It enables automatic identification of the three major plasma regions—the upstream solar wind, magnetosheath, and induced magnetosphere—and provides an efficient tool for investigating solar wind interactions and plasma dynamics under varying conditions at Mars.

\section{Methods}

\subsection{Experiment Design and Data Partitioning}
 MAVEN was inserted into an elliptical orbit around Mars with a periapsis of approximately 150 km, an apoapsis of about 6,200 km, and an orbital period of ~4.5 hours, enabling comprehensive sampling of the Martian space environment. During a typical orbit, MAVEN traverses multiple plasma regions, including the upstream solar wind (SW), the magnetosheath (MSH), and the magnetosphere (MSP).

\begin{figure}[H]
\centering
\includegraphics[width=1\textwidth]{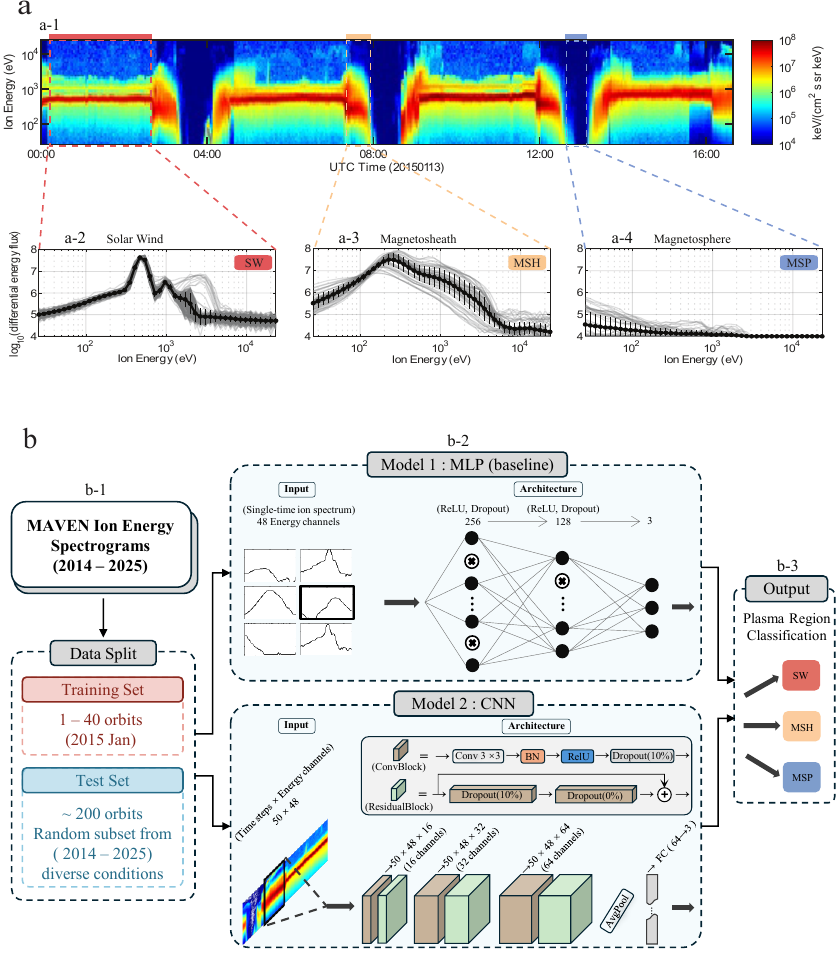}

\caption{
(a) Example MAVEN observations on 13 January 2015.
(1) Ion energy–time spectrogram with typical plasma regions, including solar wind (SW), magnetosheath (MSH), and magnetosphere (MSP).
(2–4) Ion energy spectra from the three regions: (2) SW, (3) MSH, and (4) MSP. Gray curves represent individual spectra, and the black line shows the mean spectrum after $\log_{10}$ transformation with $\pm1\sigma$ variability.
(b) Workflow of dataset construction and model evaluation. 
(1) Dataset construction: MAVEN ion energy spectrograms from 2014--2025 are used to build the dataset. A labeled training pool from January 2015 containing up to 40 orbits is used for model training, while the test set consists of about 200 orbits sampled from MAVEN observations under diverse conditions. 
(2) Model training: two models are evaluated, including a multilayer perceptron (MLP) that uses individual energy spectra as input and a convolutional neural network (CNN) that processes short time--energy spectrogram segments. 
(3) Plasma-region classification: both models output probabilities and final plasma-region labels, including SW, MSH, and MSP.
}
\label{fig1}
\end{figure}

Figure~\ref{fig1} presents representative observations from the MAVEN Solar Wind Ion Analyzer (SWIA), illustrating typical ion omnidirectional energy distributions across these different regions. In the upstream solar wind (red highlighted interval), the ion energy distribution is relatively narrow, with a pronounced peak near 1 keV corresponding to solar wind protons and a secondary peak around 2 keV associated with alpha particles. In the magnetosheath (yellow highlighted interval), ions are significantly heated and decelerated, resulting in broader and smoother energy distributions with characteristic energies of several hundred eV. Inside the magnetosphere (blue highlighted interval)—including both the induced magnetosphere and regions influenced by crustal magnetic fields—the ion flux generally decreases, and the distribution becomes more irregular compared with the SW and MSH regions. These distinct characteristics among the SW, MSH, and MSP regions in the energy distribution provide clear signatures that are well suited for automated plasma region classification using machine learning methods.

Plasma-region identification can be approached either through supervised classification or through unsupervised clustering. However, the plasma environment around Mars is highly dynamic and complex due to boundary motions, small-scale structures, and transient events. In addition, extensive prior knowledge of Martian plasma regions is available from previous observational and modeling studies. Therefore, we adopt a supervised machine learning approach, in which a subset of representative ion energy distributions is manually labeled and used to train the model to learn key features associated with different plasma regions.

For the training dataset, up to 40 representative MAVEN orbits from January 2015 are selected, as illustrated in the data partitioning scheme in Figure~\ref{fig1}b. The number of training orbits is varied from 1 to 40 to investigate how the model performance depends on the size of the training dataset. To evaluate model performance, approximately 200 MAVEN orbits from observations between 2014 and 2025 are selected as the test dataset. These test orbits span multiple years and cover a wide range of orbital geometries and solar wind conditions, providing a diverse dataset that allows us to assess the model’s generalization capability and its ability to identify plasma regions under varying conditions.

\begin{figure}[H]
\centering
\includegraphics[width=1\textwidth]{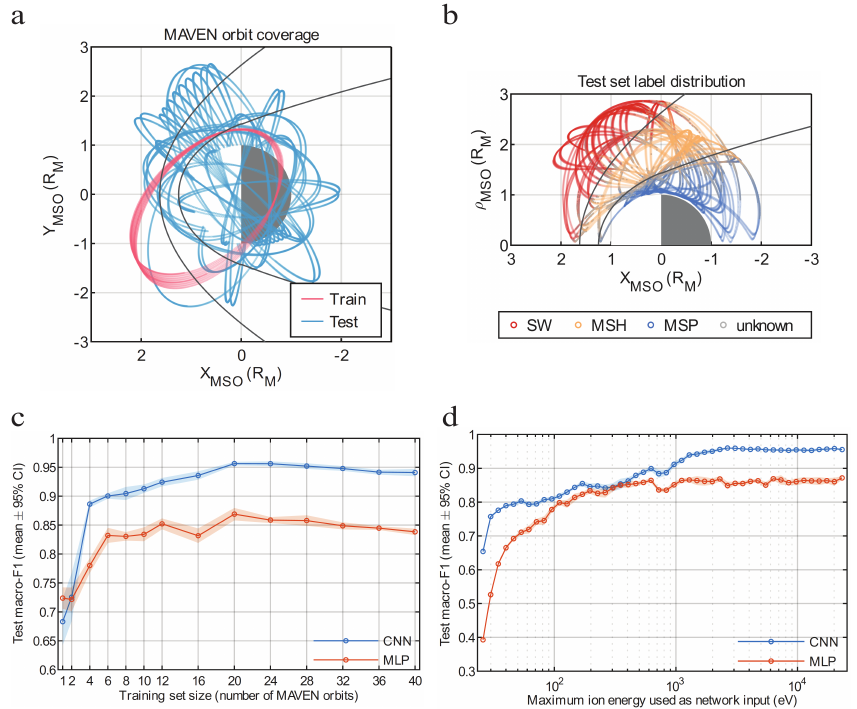}
 \caption{
(a) MAVEN orbit coverage of the training and test datasets in the MSO $X$–$Y$ plane.
(b) Spatial distribution of manually labeled plasma regions from the test dataset in the MSO $X$–$\rho$ plane, where $\rho=\sqrt{Y^2+Z^2}$.
(c) Macro-F1 score on the test dataset versus the number of training orbits for the CNN and MLP models, with training orbits progressively increased from the predefined training pool.
(d) Macro-F1 score on the test dataset as a function of the maximum ion energy used as network input (eV), with the lower bound fixed at the SWIA instrumental limit ($\sim$25 eV) and the upper bound progressively increased.
In panels (c) and (d), solid lines show the mean performance over repeated runs and shaded regions indicate the 95\% confidence interval.
}
 
\label{fig2}
\end{figure}

Figures~\ref{fig2}a and \ref{fig2}b show the spatial distribution of the training and test orbits in the $X$–$Y$ and $X$–$\rho$ planes, where $\rho=\sqrt{Y^2+Z^2}$. Here we use the Mars-Solar-Orbital (MSO) coordinate system, where $X$-axis pointing towards the Sun, $Z$-axis perpendicular to the orbital plane pointing north, and $Y$-axis completing the right-handed set. The spatial distribution of the manually labeled plasma regions is generally consistent with the empirical boundary model of \citeA{Trotignon2006}, suggesting that the labeling procedure is physically reasonable and reliable. It should be noted that ion energy distributions in the transition regions near plasma boundaries can be highly complex. To avoid introducing ambiguous samples into the training process, these cases are labeled as “unknown” and excluded from both the training and test datasets. 

\subsection{Model Architectures}

In this study, we employ two neural network architectures, as illustrated in Figure~\ref{fig1}b. 
For both models, the input consists solely of the ion energy distribution measured by the SWIA, and the output corresponds to the predicted plasma-region classes: SW, MSH, and MSP.

The first model is a feedforward multilayer perceptron (MLP) \cite{rumelhart1986learning}. 
This model extracts features from the ion energy distributions through two fully connected layers with 256 and 128 hidden units, respectively. Each layer is followed by a ReLU activation function and dropout regularization. The input to the MLP is a single-time ion omnidirectional energy spectrum with 48 energy channels. The final layer produces the class scores for the three plasma regions. Since the MLP uses only the ion energy distribution at a single time step and does not incorporate information from surrounding measurements, it primarily captures instantaneous spectral differences among plasma regions. Therefore, this model serves as a simple baseline for comparison with more advanced architectures that incorporate additional contextual information.

The second model is a convolutional neural network (CNN) \cite{lecun1998gradient}. 
In this model, ion spectra from consecutive time steps are stacked to form a two-dimensional time--energy representation. The input to the CNN consists of 50 consecutive spectra with 48 energy channels, forming a $50 \times 48$ time--energy matrix, corresponding to a 50-minute temporal window at a cadence of 1 minute. The temporal window length of 50 was chosen based on hyperparameter testing and provided stable performance. The resulting time--energy input is then processed by several convolutional stages. In each stage, a convolutional block (ConvBlock) with a $3 \times 3$ convolution layer is first applied to extract local temporal--spectral features, followed by batch normalization, ReLU activation, and dropout. A residual block with an identity skip connection \cite{he2016deep} is then applied to improve feature representation. Across convolutional stages, the number of feature channels gradually increases (16, 32, and 64), enabling the network to learn increasingly complex spectral structures. The extracted features are then aggregated through global average pooling and passed to a fully connected layer that produces the final class scores.

All models are trained using the cross-entropy loss function, and the network parameters are optimized using the AdamW optimizer \cite{loshchilov2017}. The network outputs class scores, which are converted to probabilities through a softmax function for evaluation and final classification.

\section{Results}
\subsection{Model Performance Analysis}\label{sec:model-performance}

To determine the final model configuration used in the subsequent analysis, we evaluate how training data size and the input energy range affect model performance, quantified using the macro-averaged F1 score on the test set:
\begin{equation}
\mathrm{F1}_{\mathrm{macro}}=\frac{1}{3}\sum_{k=1}^{3}\mathrm{F1}_k ,
\end{equation}
where $k$ denotes the class index and $\mathrm{F1}_k$ is the F1 score of class $k$. The F1 score for each class is defined as
\begin{equation}
\mathrm{F1}_k=\frac{2P_kR_k}{P_k+R_k},
\end{equation}
where precision $P_k$ and recall $R_k$ are given by
\begin{equation}
P_k=\frac{\mathrm{TP}_k}{\mathrm{TP}_k + \mathrm{FP}_k}, \qquad
R_k=\frac{\mathrm{TP}_k}{\mathrm{TP}_k + \mathrm{FN}_k}.
\end{equation}

Here $\mathrm{TP}_k$, $\mathrm{FP}_k$, and $\mathrm{FN}_k$ represent the numbers of true positives, false positives, and false negatives for class $k$, respectively. To evaluate the uncertainty associated with random initialization, each experiment is repeated 10 times using different random seeds. The curves in the figures show the mean performance, and the shaded regions indicate the corresponding 95\% confidence interval.

\subsubsection{Effect of Training Orbit Number}

Figure~\ref{fig2}c shows the test macro-F1 score as a function of the number of training orbits. When only a very small number of orbits are used for training, the performances of the MLP and CNN models are comparable. As the number of training orbits increases, the size of the training dataset also increases, and the performance of both models improves significantly. When the number of training orbits reaches about 6, the F1 score begins to level off, although it continues to increase gradually. The F1 scores reach their maximum when approximately 20 orbits are used for training ($\sim$0.95 for CNN and $\sim$0.87 for MLP). When the number of training orbits exceeds 20, the performance of both models begins to decrease slightly, suggesting possible overfitting. This behavior may be related to the composition of the training data: the additional orbits originate from nearby time periods and therefore represent similar orbital geometries and plasma environments, providing limited additional information. In contrast, when fewer than about 6 orbits are used, the models are likely underfitted due to the limited size of the training dataset.

Overall, the CNN consistently outperforms the MLP. This likely arises because the CNN uses the full time--energy distribution as input, allowing it to capture both distribution features and their temporal evolution. In contrast, the MLP only uses the one-dimensional energy distribution and therefore cannot exploit the additional temporal patterns.

\subsubsection{Effect of Spectral Input Range}

In addition to the training set size, we also investigate how using only a subset of the energy spectral channels affects the classification performance. In our model, the lower energy bound is fixed at the SWIA instrumental limit ($\sim$25 eV), while the upper energy bound is progressively increased to examine how the available spectral information influences the model performance. The performance is quantified using the macro-averaged F1 score, as shown in Figure~\ref{fig2}d.

As the upper energy bound increases to $\sim$100~eV, the performance of both models improves significantly. For the MLP, however, the improvement becomes minimal once the upper energy bound exceeds $\sim$400~eV, and the F1 score remains nearly constant thereafter. In contrast, the CNN performance continues to improve as the upper energy limit increases up to $\sim$1.5~keV, after which it shows little additional change. Overall, the CNN consistently outperforms the MLP, reaching an F1 score of $\sim$0.95 when the upper energy bound exceeds $\sim$1.5~keV, whereas the MLP saturates at $\sim$0.87 above $\sim$400~eV. This difference likely reflects the additional temporal and neighborhood information captured by the CNN.

Typically, an upper energy bound of $\sim$400~eV captures only part of the magnetosheath proton population and excludes the higher-energy solar wind protons and alpha particles. This may indicate that the MLP does not learn sufficiently discriminative features to clearly separate the solar wind from the magnetosheath. Consequently, its performance nearly saturates once the upper energy limit reaches $\sim$400~eV. In contrast, extending the upper energy limit to $\sim$1.5~keV includes the main solar wind ion population, providing more diagnostic information for identifying the solar wind region. This may partly explain the superior performance of the CNN.

When the input energy range extends further into higher values, the performance of both models remains nearly unchanged. This may indicate that extending the input to higher energies provides only limited additional diagnostic information for plasma-region classification. In addition, the instrumental characteristics at higher energies, where the spectral resolution and signal-to-noise ratio typically decrease, may further limit the amount of useful information gained by extending the input energy range.

\subsection{Model Predictions and Case Studies}

Based on the experiments described in Section~\ref{sec:model-performance}, the final model configuration adopts 20 training orbits and uses the full SWIA energy channels as the neural network input. Under this optimal setup, we compare the predictions of the MLP and CNN models using both the test dataset and all available MAVEN observations.

\subsubsection{Spatial Classification Results}

\begin{figure}[H]
\centering
\includegraphics[width=1\textwidth]{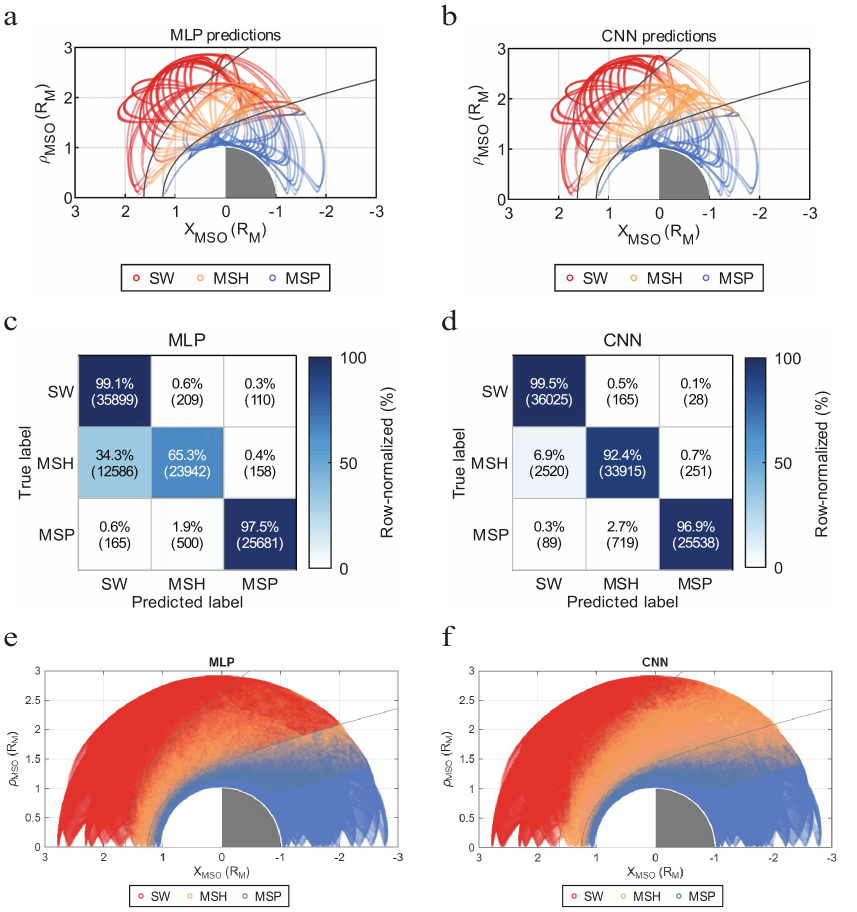}

\caption{
Model predictions and classification performance.
Plasma-region predictions on the test dataset in the MSO $X$--$\rho$ projection: (a) MLP and (b) CNN.
Normalized confusion matrices on the test dataset: (c) MLP and (d) CNN.
Plasma-region classification maps in the MSO $X$--$\rho$ projection using MAVEN observations from 2014 to 2025 across all available orbits, showing high-confidence predictions from the optimal models: (e) MLP and (f) CNN.
}
\label{fig3}
\end{figure}

We first apply both models to the test dataset and evaluate their predictions and performance, as shown in Figures~\ref{fig3}a--\ref{fig3}d. Comparing Figures~\ref{fig3}a and~\ref{fig3}b with the ground truth distribution shown in Figure~\ref{fig2}b, both the MLP and CNN predictions generally reproduce the expected large-scale plasma structure around Mars and agree well with the nominal bow shock and MPB models of \citeA{Trotignon2006}. However, noticeable differences exist between the two models. In the MLP results, many data points in the nominal MSH, particularly in the flank regions, are classified as SW, whereas this issue is largely absent in the CNN results.
This behavior indicates that the MLP has more difficulty distinguishing between SW and MSH than the CNN. This is also evident in Figure~\ref{fig3}c, which shows that in the MLP model 34.3\% of MSH data points are misclassified as SW, and only 65.3\% of MSH samples are correctly identified. In contrast, as shown in Figure~\ref{fig3}d, the CNN misclassifies only 6.9\% of MSH data points as SW, demonstrating a much stronger ability to distinguish between SW and MSH. For distinguishing between SW and MSP, as well as between MSP and MSH, both models perform well. As shown in Figures~\ref{fig3}c--\ref{fig3}d, the misclassification rate is less than 3\%.

We then apply both models to all available MAVEN observations from 2014 to 2025, as shown in Figures~\ref{fig3}e--\ref{fig3}f. In these full-mission maps, only predictions with a maximum class probability greater than 0.8 are shown to reduce low-confidence classifications. In the dayside region, the predictions from both the MLP and CNN models are generally consistent with the nominal boundaries predicted by \citeA{Trotignon2006}. However, noticeable differences appear in the nightside flank regions. In particular, the MLP model classifies many outer MSH data points as SW, whereas this issue is much less pronounced in the CNN results. In addition, in the nightside flank region, both the CNN and MLP models predict the boundary between MSH and MSP to be located significantly farther inward than the MPB reported by~\citeA{Trotignon2006}. This difference likely arises because the MPB in~\citeA{Trotignon2006} was identified using magnetic field measurements, whereas our models rely solely on ion measurements. Therefore, our method is more likely to identify the ion composition boundary (ICB). Although the MPB and ICB are generally located close to each other, previous studies have shown that the ICB is typically situated slightly inward of the MPB \cite{Matsunaga2017}.

\subsubsection{Example Orbits}

\begin{figure}[H]
\centering
\includegraphics[width=1\textwidth]{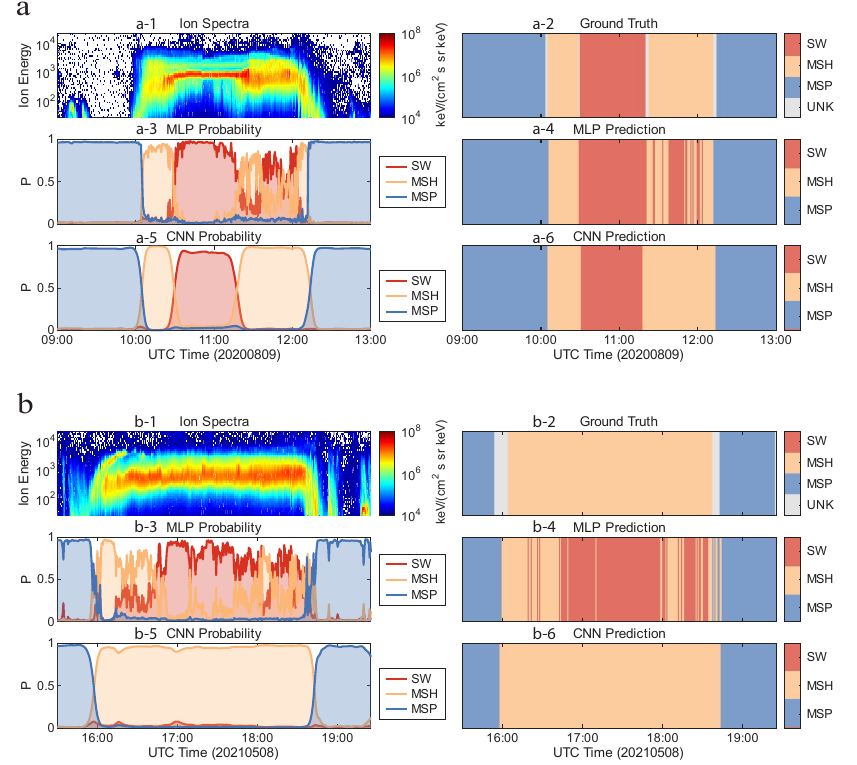}

\caption{
Predictions for two MAVEN orbital segments selected from the test dataset.
(a) Orbital segment that crosses all three plasma regions (SW, MSH, and MSP).
(b) Orbital segment that remains within the MSH and MSP without entering the SW.
For each case, the panels show:
(1) ion energy spectrogram,
(2) ground-truth plasma-region labels,
(3) MLP class probabilities,
(4) MLP predictions,
(5) CNN class probabilities,
(6) CNN predictions.
}
\label{fig4}
\end{figure}

Two representative MAVEN orbits are further evaluated for the performance of the models and compared against the predictions from the ground truth in detail (Figure~\ref{fig4}). The first example corresponds to an orbit that crosses all three plasma regions (SW, MSH, and MSP), whereas the second example corresponds to an orbit that does not enter the SW region and only samples the MSH and MSP regions.

In the first case (Figure~\ref{fig4}a), MAVEN was initially located in the MSP between 09:00 and 10:05. During 10:05–10:30, the spacecraft crossed into the MSH region, where hot ions were detected. Around 10:30, MAVEN entered the SW region and remained there until approximately 11:15, after which it re-entered the MSH. At about 12:10, MAVEN crossed back into the MSP region. In the boundary and transition regions, it is difficult to determine the plasma region precisely because the ion properties are mixed. Therefore, these intervals are labeled as unknown (“UNK”) in the ground truth. Overall, both the MLP and CNN models perform well for most of the orbit. However, differences appear during the second MSH interval, where the MLP misclassifies many data points as SW, which is inconsistent with the ground truth. In contrast, the CNN provides more accurate predictions that remain consistent with the labeled regions.

In the second case (Figure~\ref{fig4}b), MAVEN remained in the MSH region between 16:00 and 18:40, characterized by the detection of hot ions. During the remaining time, the spacecraft was located in the MSP region, with no bow shock crossings or solar wind signatures observed. In this case, the CNN predictions are almost entirely consistent with the ground truth. In contrast, the performance of the MLP deteriorates further, incorrectly classifying many data points as SW. This also suggests that the MLP struggles to distinguish between the MSH and SW regions, whereas the CNN demonstrates a much better capability in separating these plasma regimes.

\section{Conclusions and Discussions}

In this study, we developed a machine learning–based approach to automatically classify plasma regions at Mars using MAVEN ion omnidirectional energy spectra. The method identifies the three primary plasma regions—solar wind (SW), magnetosheath (MSH), and magnetosphere (MSP). Unlike previous approaches that rely on multiple plasma parameters or complex physical diagnostics, our method uses only ion energy spectra as input and allows machine learning models to directly learn the spectral features that distinguish different plasma regions.

The results show that ion energy spectra alone contain sufficient physical information to distinguish between different plasma regions. Even when trained using a limited number of labeled orbits, the models achieve high classification performance and generalize well to observations from different years and orbital conditions. In particular, the CNN, which exploits the short-timescale temporal continuity in the spectra, demonstrates more stable and accurate performance than the MLP, which relies only on single-time spectra as input. This advantage is especially evident in distinguishing regions with similar spectral characteristics, such as the solar wind and magnetosheath.

Under the optimal model configuration, the CNN achieves a macro-averaged F1 score of approximately 95\% on the test dataset. To further assess its robustness, we applied the trained model to all available MAVEN orbits from 2014 to 2025. The resulting classifications are largely consistent with traditional empirical models, demonstrating that the model can provide reliable identification of plasma regions at Mars over a wide range of observing conditions.

The model also has clear practical value for MAVEN data analysis. It enables automatic identification of plasma regions and the associated boundaries, which can help determine the upstream solar wind conditions associated with specific events. This makes it a useful tool for event studies, large statistical analyses, and rapid determination of solar wind context in MAVEN observations.

Overall, this study demonstrates that a CNN-based approach can efficiently and accurately identify plasma regions at Mars. The method provides a practical framework for investigating solar wind–Mars interactions and related plasma processes. Because it does not require complex parameter derivation or extensive data preprocessing, the approach is computationally efficient and can be readily applied to observations from future planetary missions, such as ESCAPADE, for plasma environment identification.

\section{Figures and Figure Captions}

\section*{Data Availability Statement}
MAVEN magnetic field data are publicly available through the MAVEN Science Data Center at \url{https://lasp.colorado.edu/maven/sdc/public/data/sci/mag/l2/} \cite{Connerney2015}. MAVEN SWIA data are publicly available through the MAVEN Science Data Center at \url{https://lasp.colorado.edu/maven/sdc/public/data/sci/swi/l2/} \cite{Halekas2017}.

\acknowledgments
This work was partially supported by NASA grants 80NSSC23K1125 and 80NSSC24K1843, the Alfred P. Sloan Research Fellowship, and the IBM Einstein Fellow Fund at the Institute for Advanced Study, Princeton.


\begin{thebibliography}{}

\bibitem [\protect \citeauthoryear {%
Cheng%
, Achilleos%
\BCBL {}\ \BBA {} Smith%
}{%
Cheng%
\ \protect \BOthers {.}}{%
{\protect \APACyear {2022}}%
}]{%
Cheng2022}
\APACinsertmetastar {%
Cheng2022}%
\begin{APACrefauthors}%
Cheng, I\BPBI K.%
, Achilleos, N.%
\BCBL {}\ \BBA {} Smith, A.%
\end{APACrefauthors}%
\unskip\
\newblock
\APACrefYearMonthDay{2022}{}{}.
\newblock
{\BBOQ}\APACrefatitle {Automated bow shock and magnetopause boundary detection
  with {Cassini} using threshold and deep learning methods} {Automated bow
  shock and magnetopause boundary detection with {Cassini} using threshold and
  deep learning methods}.{\BBCQ}
\newblock
\APACjournalVolNumPages{Frontiers in Astronomy and Space
  Sciences}{9}{}{1016453}.
\newblock
\begin{APACrefDOI} \doi{10.3389/fspas.2022.1016453} \end{APACrefDOI}
\PrintBackRefs{\CurrentBib}

\bibitem [\protect \citeauthoryear {%
{Cheng}%
\ \protect \BOthers {.}}{%
{Cheng}%
\ \protect \BOthers {.}}{%
{\protect \APACyear {2025}}%
}]{%
Cheng2025}
\APACinsertmetastar {%
Cheng2025}%
\begin{APACrefauthors}%
{Cheng}, Z.%
, {Zhang}, C.%
, {Dong}, C.%
, {Zhou}, H.%
, {Gao}, J.%
, {Tadlock}, A.%
\BDBL {}{Wang}, L.%
\end{APACrefauthors}%
\unskip\
\newblock
\APACrefYearMonthDay{2025}{{\APACmonth{12}}}{}.
\newblock
{\BBOQ}\APACrefatitle {{Revisiting Mars' Induced Magnetic Field and Clock Angle
  Departures Under Real-Time Upstream Solar Wind Conditions}} {{Revisiting
  Mars' Induced Magnetic Field and Clock Angle Departures Under Real-Time
  Upstream Solar Wind Conditions}}.{\BBCQ}
\newblock
\APACjournalVolNumPages{Journal of Geophysical Research (Space
  Physics)}{130}{12}{e2025JA034688}.
\newblock
\begin{APACrefDOI} \doi{10.1029/2025JA034688} \end{APACrefDOI}
\PrintBackRefs{\CurrentBib}

\bibitem [\protect \citeauthoryear {%
Connerney%
\ \protect \BOthers {.}}{%
Connerney%
\ \protect \BOthers {.}}{%
{\protect \APACyear {2015}}%
}]{%
Connerney2015}
\APACinsertmetastar {%
Connerney2015}%
\begin{APACrefauthors}%
Connerney, J\BPBI E\BPBI P.%
, Espley, J.%
, Lawton, P.%
, Murphy, S.%
, Odom, J.%
, Oliversen, R.%
\BCBL {}\ \BBA {} Sheppard, D.%
\end{APACrefauthors}%
\unskip\
\newblock
\APACrefYearMonthDay{2015}{}{}.
\newblock
{\BBOQ}\APACrefatitle {The {MAVEN} Magnetic Field Investigation} {The {MAVEN}
  magnetic field investigation}.{\BBCQ}
\newblock
\APACjournalVolNumPages{Space Science Reviews}{195}{1--4}{257--291}.
\newblock
\begin{APACrefDOI} \doi{10.1007/s11214-015-0169-4} \end{APACrefDOI}
\PrintBackRefs{\CurrentBib}

\bibitem [\protect \citeauthoryear {%
Dong%
\ \protect \BOthers {.}}{%
Dong%
\ \protect \BOthers {.}}{%
{\protect \APACyear {2014}}%
}]{%
Dong2014}
\APACinsertmetastar {%
Dong2014}%
\begin{APACrefauthors}%
Dong, C.%
, Bougher, S\BPBI W.%
, Ma, Y.%
, Toth, G.%
, Nagy, A\BPBI F.%
\BCBL {}\ \BBA {} Najib, D.%
\end{APACrefauthors}%
\unskip\
\newblock
\APACrefYearMonthDay{2014}{}{}.
\newblock
{\BBOQ}\APACrefatitle {Solar wind interaction with Mars upper atmosphere}
  {Solar wind interaction with mars upper atmosphere}.{\BBCQ}
\newblock
\APACjournalVolNumPages{Geophysical Research Letters}{41}{}{2708--2715}.
\newblock
\begin{APACrefDOI} \doi{10.1002/2014GL059515} \end{APACrefDOI}
\PrintBackRefs{\CurrentBib}

\bibitem [\protect \citeauthoryear {%
Dong%
\ \protect \BOthers {.}}{%
Dong%
\ \protect \BOthers {.}}{%
{\protect \APACyear {2015}}%
}]{%
Dong2015}
\APACinsertmetastar {%
Dong2015}%
\begin{APACrefauthors}%
Dong, C.%
\BCBT {}\ \BOthersPeriod {.}
\end{APACrefauthors}%
\unskip\
\newblock
\APACrefYearMonthDay{2015}{}{}.
\newblock
{\BBOQ}\APACrefatitle {Solar wind interaction with the Martian upper
  atmosphere} {Solar wind interaction with the martian upper
  atmosphere}.{\BBCQ}
\newblock
\APACjournalVolNumPages{Journal of Geophysical Research: Space
  Physics}{120}{}{7857--7872}.
\newblock
\begin{APACrefDOI} \doi{10.1002/2015JA020990} \end{APACrefDOI}
\PrintBackRefs{\CurrentBib}

\bibitem [\protect \citeauthoryear {%
Dong%
\ \protect \BOthers {.}}{%
Dong%
\ \protect \BOthers {.}}{%
{\protect \APACyear {2018}}%
{\protect \APACexlab {{\protect \BCnt {1}}}}}]{%
Dong2018a}
\APACinsertmetastar {%
Dong2018a}%
\begin{APACrefauthors}%
Dong, C.%
\BCBT {}\ \BOthersPeriod {.}
\end{APACrefauthors}%
\unskip\
\newblock
\APACrefYearMonthDay{2018{\protect \BCnt {1}}}{}{}.
\newblock
{\BBOQ}\APACrefatitle {Modeling Martian Atmospheric Losses over Time} {Modeling
  martian atmospheric losses over time}.{\BBCQ}
\newblock
\APACjournalVolNumPages{The Astrophysical Journal}{859}{}{}.
\newblock
\begin{APACrefDOI} \doi{10.3847/2041-8213/aac489} \end{APACrefDOI}
\PrintBackRefs{\CurrentBib}

\bibitem [\protect \citeauthoryear {%
Dong%
\ \protect \BOthers {.}}{%
Dong%
\ \protect \BOthers {.}}{%
{\protect \APACyear {2018}}%
{\protect \APACexlab {{\protect \BCnt {2}}}}}]{%
Dong2018b}
\APACinsertmetastar {%
Dong2018b}%
\begin{APACrefauthors}%
Dong, C.%
\BCBT {}\ \BOthersPeriod {.}
\end{APACrefauthors}%
\unskip\
\newblock
\APACrefYearMonthDay{2018{\protect \BCnt {2}}}{}{}.
\newblock
{\BBOQ}\APACrefatitle {Solar wind interaction with the Martian upper
  atmosphere} {Solar wind interaction with the martian upper
  atmosphere}.{\BBCQ}
\newblock
\APACjournalVolNumPages{Journal of Geophysical Research: Space
  Physics}{123}{}{6639--6654}.
\newblock
\begin{APACrefDOI} \doi{10.1029/2018JA025543} \end{APACrefDOI}
\PrintBackRefs{\CurrentBib}

\bibitem [\protect \citeauthoryear {%
Dubinin%
\ \protect \BOthers {.}}{%
Dubinin%
\ \protect \BOthers {.}}{%
{\protect \APACyear {2017}}%
}]{%
Dubinin2017}
\APACinsertmetastar {%
Dubinin2017}%
\begin{APACrefauthors}%
Dubinin, E.%
\BCBT {}\ \BOthersPeriod {.}
\end{APACrefauthors}%
\unskip\
\newblock
\APACrefYearMonthDay{2017}{}{}.
\newblock
{\BBOQ}\APACrefatitle {The Effect of Solar Wind Variations on the Escape of
  Oxygen Ions From Mars} {The effect of solar wind variations on the escape of
  oxygen ions from mars}.{\BBCQ}
\newblock
\APACjournalVolNumPages{Journal of Geophysical Research: Space
  Physics}{122}{}{}.
\newblock
\begin{APACrefDOI} \doi{10.1002/2017JA024741} \end{APACrefDOI}
\PrintBackRefs{\CurrentBib}

\bibitem [\protect \citeauthoryear {%
Halekas%
\ \protect \BOthers {.}}{%
Halekas%
\ \protect \BOthers {.}}{%
{\protect \APACyear {2017}}%
}]{%
Halekas2017}
\APACinsertmetastar {%
Halekas2017}%
\begin{APACrefauthors}%
Halekas, J\BPBI S.%
\BCBT {}\ \BOthersPeriod {.}
\end{APACrefauthors}%
\unskip\
\newblock
\APACrefYearMonthDay{2017}{}{}.
\newblock
{\BBOQ}\APACrefatitle {Structure, dynamics, and seasonal variability of the
  Mars-solar wind interaction} {Structure, dynamics, and seasonal variability
  of the mars-solar wind interaction}.{\BBCQ}
\newblock
\APACjournalVolNumPages{Journal of Geophysical Research: Space
  Physics}{122}{}{547--578}.
\PrintBackRefs{\CurrentBib}

\bibitem [\protect \citeauthoryear {%
Halekas%
\ \protect \BOthers {.}}{%
Halekas%
\ \protect \BOthers {.}}{%
{\protect \APACyear {2021}}%
}]{%
Halekas2021}
\APACinsertmetastar {%
Halekas2021}%
\begin{APACrefauthors}%
Halekas, J\BPBI S.%
\BCBT {}\ \BOthersPeriod {.}
\end{APACrefauthors}%
\unskip\
\newblock
\APACrefYearMonthDay{2021}{}{}.
\newblock
{\BBOQ}\APACrefatitle {Induced Magnetospheres} {Induced magnetospheres}.{\BBCQ}
\newblock
\APACjournalVolNumPages{Magnetospheres in the Solar System}{}{}{}.
\newblock
\begin{APACrefDOI} \doi{10.1002/9781119815624.ch25} \end{APACrefDOI}
\PrintBackRefs{\CurrentBib}

\bibitem [\protect \citeauthoryear {%
He%
, Zhang%
, Ren%
\BCBL {}\ \BBA {} Sun%
}{%
He%
\ \protect \BOthers {.}}{%
{\protect \APACyear {2016}}%
}]{%
he2016deep}
\APACinsertmetastar {%
he2016deep}%
\begin{APACrefauthors}%
He, K.%
, Zhang, X.%
, Ren, S.%
\BCBL {}\ \BBA {} Sun, J.%
\end{APACrefauthors}%
\unskip\
\newblock
\APACrefYearMonthDay{2016}{}{}.
\newblock
{\BBOQ}\APACrefatitle {Deep residual learning for image recognition} {Deep
  residual learning for image recognition}.{\BBCQ}
\newblock
\BIn{} \APACrefbtitle {Proceedings of the IEEE conference on computer vision
  and pattern recognition} {Proceedings of the ieee conference on computer
  vision and pattern recognition}\ (\BPGS\ 770--778).
\PrintBackRefs{\CurrentBib}

\bibitem [\protect \citeauthoryear {%
Hollman%
\ \protect \BOthers {.}}{%
Hollman%
\ \protect \BOthers {.}}{%
{\protect \APACyear {2026}}%
}]{%
Hollman2026}
\APACinsertmetastar {%
Hollman2026}%
\begin{APACrefauthors}%
Hollman, D\BPBI M.%
, Jackman, C\BPBI M.%
, Domijan, K.%
, Bowers, C\BPBI F.%
, Walker, S\BPBI J.%
, Rutala, M\BPBI J.%
\BCBL {}\ \BBA {} Fogg, A\BPBI R.%
\end{APACrefauthors}%
\unskip\
\newblock
\APACrefYearMonthDay{2026}{}{}.
\newblock
{\BBOQ}\APACrefatitle {Identifying MESSENGER magnetospheric boundary crossings
  using a random forest region classifier} {Identifying messenger
  magnetospheric boundary crossings using a random forest region
  classifier}.{\BBCQ}
\newblock
\APACjournalVolNumPages{Journal of Geophysical Research: Machine Learning and
  Computation}{3}{1}{e2025JH000921}.
\newblock
\begin{APACrefDOI} \doi{10.1029/2025JH000921} \end{APACrefDOI}
\PrintBackRefs{\CurrentBib}

\bibitem [\protect \citeauthoryear {%
Inui%
\ \protect \BOthers {.}}{%
Inui%
\ \protect \BOthers {.}}{%
{\protect \APACyear {2018}}%
}]{%
Inui2018}
\APACinsertmetastar {%
Inui2018}%
\begin{APACrefauthors}%
Inui, S.%
\BCBT {}\ \BOthersPeriod {.}
\end{APACrefauthors}%
\unskip\
\newblock
\APACrefYearMonthDay{2018}{}{}.
\newblock
{\BBOQ}\APACrefatitle {Cold Dense Ion Outflow Observed in the Martian-Induced
  Magnetotail} {Cold dense ion outflow observed in the martian-induced
  magnetotail}.{\BBCQ}
\newblock
\APACjournalVolNumPages{Geophysical Research Letters}{45}{}{}.
\newblock
\begin{APACrefDOI} \doi{10.1029/2018GL077584} \end{APACrefDOI}
\PrintBackRefs{\CurrentBib}

\bibitem [\protect \citeauthoryear {%
Jakosky%
\ \protect \BOthers {.}}{%
Jakosky%
\ \protect \BOthers {.}}{%
{\protect \APACyear {2015}}%
}]{%
Jakosky2015}
\APACinsertmetastar {%
Jakosky2015}%
\begin{APACrefauthors}%
Jakosky, B\BPBI M.%
\BCBT {}\ \BOthersPeriod {.}
\end{APACrefauthors}%
\unskip\
\newblock
\APACrefYearMonthDay{2015}{}{}.
\newblock
{\BBOQ}\APACrefatitle {The Mars Atmosphere and Volatile Evolution Mission} {The
  mars atmosphere and volatile evolution mission}.{\BBCQ}
\newblock
\APACjournalVolNumPages{Space Science Reviews}{195}{}{3--48}.
\newblock
\begin{APACrefDOI} \doi{10.1007/s11214-015-0139-x} \end{APACrefDOI}
\PrintBackRefs{\CurrentBib}

\bibitem [\protect \citeauthoryear {%
Jakosky%
\ \protect \BOthers {.}}{%
Jakosky%
\ \protect \BOthers {.}}{%
{\protect \APACyear {2018}}%
}]{%
Jakosky2018}
\APACinsertmetastar {%
Jakosky2018}%
\begin{APACrefauthors}%
Jakosky, B\BPBI M.%
\BCBT {}\ \BOthersPeriod {.}
\end{APACrefauthors}%
\unskip\
\newblock
\APACrefYearMonthDay{2018}{}{}.
\newblock
{\BBOQ}\APACrefatitle {Loss of the Martian atmosphere to space} {Loss of the
  martian atmosphere to space}.{\BBCQ}
\newblock
\APACjournalVolNumPages{Icarus}{315}{}{146--157}.
\newblock
\begin{APACrefDOI} \doi{10.1016/j.icarus.2018.05.030} \end{APACrefDOI}
\PrintBackRefs{\CurrentBib}

\bibitem [\protect \citeauthoryear {%
LeCun%
, Bottou%
, Bengio%
\BCBL {}\ \BBA {} Haffner%
}{%
LeCun%
\ \protect \BOthers {.}}{%
{\protect \APACyear {1998}}%
}]{%
lecun1998gradient}
\APACinsertmetastar {%
lecun1998gradient}%
\begin{APACrefauthors}%
LeCun, Y.%
, Bottou, L.%
, Bengio, Y.%
\BCBL {}\ \BBA {} Haffner, P.%
\end{APACrefauthors}%
\unskip\
\newblock
\APACrefYearMonthDay{1998}{}{}.
\newblock
{\BBOQ}\APACrefatitle {Gradient-based learning applied to document recognition}
  {Gradient-based learning applied to document recognition}.{\BBCQ}
\newblock
\APACjournalVolNumPages{Proceedings of the IEEE}{86}{11}{2278--2324}.
\newblock
\begin{APACrefDOI} \doi{10.1109/5.726791} \end{APACrefDOI}
\PrintBackRefs{\CurrentBib}

\bibitem [\protect \citeauthoryear {%
Lillis%
\ \protect \BOthers {.}}{%
Lillis%
\ \protect \BOthers {.}}{%
{\protect \APACyear {2015}}%
}]{%
Lillis2015}
\APACinsertmetastar {%
Lillis2015}%
\begin{APACrefauthors}%
Lillis, R\BPBI J.%
\BCBT {}\ \BOthersPeriod {.}
\end{APACrefauthors}%
\unskip\
\newblock
\APACrefYearMonthDay{2015}{}{}.
\newblock
{\BBOQ}\APACrefatitle {Characterizing Atmospheric Escape from Mars Today and
  Through Time} {Characterizing atmospheric escape from mars today and through
  time}.{\BBCQ}
\newblock
\APACjournalVolNumPages{Space Science Reviews}{195}{}{357--422}.
\newblock
\begin{APACrefDOI} \doi{10.1007/s11214-015-0165-8} \end{APACrefDOI}
\PrintBackRefs{\CurrentBib}

\bibitem [\protect \citeauthoryear {%
Linzmayer%
, Nemec%
, Nemecek%
\BCBL {}\ \BBA {} Safrankova%
}{%
Linzmayer%
\ \protect \BOthers {.}}{%
{\protect \APACyear {2024}}%
}]{%
Linzmayer2024}
\APACinsertmetastar {%
Linzmayer2024}%
\begin{APACrefauthors}%
Linzmayer, V.%
, Nemec, F.%
, Nemecek, Z.%
\BCBL {}\ \BBA {} Safrankova, J.%
\end{APACrefauthors}%
\unskip\
\newblock
\APACrefYearMonthDay{2024}{}{}.
\newblock
{\BBOQ}\APACrefatitle {Martian bow shock and magnetic pileup boundary models
  based on machine learning} {Martian bow shock and magnetic pileup boundary
  models based on machine learning}.{\BBCQ}
\newblock
\APACjournalVolNumPages{Advances in Space Research}{73}{}{}.
\newblock
\begin{APACrefDOI} \doi{10.1016/j.asr.2024.03.030} \end{APACrefDOI}
\PrintBackRefs{\CurrentBib}

\bibitem [\protect \citeauthoryear {%
Loshchilov%
\ \BBA {} Hutter%
}{%
Loshchilov%
\ \BBA {} Hutter%
}{%
{\protect \APACyear {2017}}%
}]{%
loshchilov2017}
\APACinsertmetastar {%
loshchilov2017}%
\begin{APACrefauthors}%
Loshchilov, I.%
\BCBT {}\ \BBA {} Hutter, F.%
\end{APACrefauthors}%
\unskip\
\newblock
\APACrefYearMonthDay{2017}{}{}.
\newblock
{\BBOQ}\APACrefatitle {Decoupled weight decay regularization} {Decoupled weight
  decay regularization}.{\BBCQ}
\newblock
\APACjournalVolNumPages{arXiv preprint arXiv:1711.05101}{}{}{}.
\PrintBackRefs{\CurrentBib}

\bibitem [\protect \citeauthoryear {%
Luhmann%
, Ledvina%
\BCBL {}\ \BBA {} Russell%
}{%
Luhmann%
\ \protect \BOthers {.}}{%
{\protect \APACyear {2004}}%
}]{%
Luhmann2004}
\APACinsertmetastar {%
Luhmann2004}%
\begin{APACrefauthors}%
Luhmann, J\BPBI G.%
, Ledvina, S\BPBI A.%
\BCBL {}\ \BBA {} Russell, C\BPBI T.%
\end{APACrefauthors}%
\unskip\
\newblock
\APACrefYearMonthDay{2004}{}{}.
\newblock
{\BBOQ}\APACrefatitle {Induced magnetospheres} {Induced magnetospheres}.{\BBCQ}
\newblock
\APACjournalVolNumPages{Advances in Space Research}{33}{}{1905--1912}.
\newblock
\begin{APACrefDOI} \doi{10.1016/j.asr.2003.03.031} \end{APACrefDOI}
\PrintBackRefs{\CurrentBib}

\bibitem [\protect \citeauthoryear {%
Matsunaga%
\ \protect \BOthers {.}}{%
Matsunaga%
\ \protect \BOthers {.}}{%
{\protect \APACyear {2017}}%
}]{%
Matsunaga2017}
\APACinsertmetastar {%
Matsunaga2017}%
\begin{APACrefauthors}%
Matsunaga, K.%
, Terada, N.%
, Harada, Y.%
, Halekas, J\BPBI S.%
, Brain, D\BPBI A.%
, McFadden, J\BPBI P.%
\BCBL {}\ \BBA {} Jakosky, B\BPBI M.%
\end{APACrefauthors}%
\unskip\
\newblock
\APACrefYearMonthDay{2017}{}{}.
\newblock
{\BBOQ}\APACrefatitle {Statistical Study of Relations Between the Induced
  Magnetosphere, Ion Composition, and Pressure Balance Boundaries Around Mars
  Based on MAVEN Observations} {Statistical study of relations between the
  induced magnetosphere, ion composition, and pressure balance boundaries
  around mars based on maven observations}.{\BBCQ}
\newblock
\APACjournalVolNumPages{Journal of Geophysical Research: Space
  Physics}{122}{9}{9723--9737}.
\newblock
\begin{APACrefDOI} \doi{10.1002/2017JA024217} \end{APACrefDOI}
\PrintBackRefs{\CurrentBib}

\bibitem [\protect \citeauthoryear {%
Nagy%
\ \protect \BOthers {.}}{%
Nagy%
\ \protect \BOthers {.}}{%
{\protect \APACyear {2004}}%
}]{%
Nagy2004}
\APACinsertmetastar {%
Nagy2004}%
\begin{APACrefauthors}%
Nagy, A\BPBI F.%
\BCBT {}\ \BOthersPeriod {.}
\end{APACrefauthors}%
\unskip\
\newblock
\APACrefYearMonthDay{2004}{}{}.
\newblock
{\BBOQ}\APACrefatitle {The plasma environment of Mars} {The plasma environment
  of mars}.{\BBCQ}
\newblock
\APACjournalVolNumPages{Space Science Reviews}{111}{}{33--114}.
\PrintBackRefs{\CurrentBib}

\bibitem [\protect \citeauthoryear {%
Nemec%
\ \protect \BOthers {.}}{%
Nemec%
\ \protect \BOthers {.}}{%
{\protect \APACyear {2020}}%
}]{%
Nemec2020}
\APACinsertmetastar {%
Nemec2020}%
\begin{APACrefauthors}%
Nemec, F.%
\BCBT {}\ \BOthersPeriod {.}
\end{APACrefauthors}%
\unskip\
\newblock
\APACrefYearMonthDay{2020}{}{}.
\newblock
{\BBOQ}\APACrefatitle {Martian Bow Shock and Magnetic Pileup Boundary Models
  Based on an Automated Region Identification} {Martian bow shock and magnetic
  pileup boundary models based on an automated region identification}.{\BBCQ}
\newblock
\APACjournalVolNumPages{Journal of Geophysical Research: Space
  Physics}{125}{}{}.
\newblock
\begin{APACrefDOI} \doi{10.1029/2020JA028509} \end{APACrefDOI}
\PrintBackRefs{\CurrentBib}

\bibitem [\protect \citeauthoryear {%
Olshevsky%
\ \protect \BOthers {.}}{%
Olshevsky%
\ \protect \BOthers {.}}{%
{\protect \APACyear {2021}}%
}]{%
Olshevsky2021}
\APACinsertmetastar {%
Olshevsky2021}%
\begin{APACrefauthors}%
Olshevsky, V.%
, Khotyaintsev, Y\BPBI V.%
, Lalti, A.%
, Divin, A.%
, Delzanno, G\BPBI L.%
, Anderz{\'e}n, S.%
\BDBL {}Markidis, S.%
\end{APACrefauthors}%
\unskip\
\newblock
\APACrefYearMonthDay{2021}{}{}.
\newblock
{\BBOQ}\APACrefatitle {Automated Classification of Plasma Regions Using 3D
  Particle Energy Distributions} {Automated classification of plasma regions
  using 3d particle energy distributions}.{\BBCQ}
\newblock
\APACjournalVolNumPages{Journal of Geophysical Research: Space
  Physics}{126}{10}{e2021JA029620}.
\newblock
\begin{APACrefDOI} \doi{10.1029/2021JA029620} \end{APACrefDOI}
\PrintBackRefs{\CurrentBib}

\bibitem [\protect \citeauthoryear {%
Ramstad%
\ \protect \BOthers {.}}{%
Ramstad%
\ \protect \BOthers {.}}{%
{\protect \APACyear {2018}}%
}]{%
Ramstad2018}
\APACinsertmetastar {%
Ramstad2018}%
\begin{APACrefauthors}%
Ramstad, R.%
\BCBT {}\ \BOthersPeriod {.}
\end{APACrefauthors}%
\unskip\
\newblock
\APACrefYearMonthDay{2018}{}{}.
\newblock
{\BBOQ}\APACrefatitle {Ion Escape From Mars Through Time} {Ion escape from mars
  through time}.{\BBCQ}
\newblock
\APACjournalVolNumPages{Journal of Geophysical Research:
  Planets}{123}{}{3051--3060}.
\newblock
\begin{APACrefDOI} \doi{10.1029/2018JE005727} \end{APACrefDOI}
\PrintBackRefs{\CurrentBib}

\bibitem [\protect \citeauthoryear {%
Ramstad%
\ \protect \BOthers {.}}{%
Ramstad%
\ \protect \BOthers {.}}{%
{\protect \APACyear {2020}}%
}]{%
Ramstad2020}
\APACinsertmetastar {%
Ramstad2020}%
\begin{APACrefauthors}%
Ramstad, R.%
\BCBT {}\ \BOthersPeriod {.}
\end{APACrefauthors}%
\unskip\
\newblock
\APACrefYearMonthDay{2020}{}{}.
\newblock
{\BBOQ}\APACrefatitle {The global current systems of the Martian induced
  magnetosphere} {The global current systems of the martian induced
  magnetosphere}.{\BBCQ}
\newblock
\APACjournalVolNumPages{Nature Astronomy}{4}{}{979--985}.
\newblock
\begin{APACrefDOI} \doi{10.1038/s41550-020-1099-y} \end{APACrefDOI}
\PrintBackRefs{\CurrentBib}

\bibitem [\protect \citeauthoryear {%
Rumelhart%
, Hinton%
\BCBL {}\ \BBA {} Williams%
}{%
Rumelhart%
\ \protect \BOthers {.}}{%
{\protect \APACyear {1986}}%
}]{%
rumelhart1986learning}
\APACinsertmetastar {%
rumelhart1986learning}%
\begin{APACrefauthors}%
Rumelhart, D\BPBI E.%
, Hinton, G\BPBI E.%
\BCBL {}\ \BBA {} Williams, R\BPBI J.%
\end{APACrefauthors}%
\unskip\
\newblock
\APACrefYearMonthDay{1986}{}{}.
\newblock
{\BBOQ}\APACrefatitle {Learning representations by back-propagating errors}
  {Learning representations by back-propagating errors}.{\BBCQ}
\newblock
\APACjournalVolNumPages{Nature}{323}{6088}{533--536}.
\newblock
\begin{APACrefDOI} \doi{10.1038/323533a0} \end{APACrefDOI}
\PrintBackRefs{\CurrentBib}

\bibitem [\protect \citeauthoryear {%
Trotignon%
, Mazelle%
, Bertucci%
\BCBL {}\ \BBA {} Acu{\~n}a%
}{%
Trotignon%
\ \protect \BOthers {.}}{%
{\protect \APACyear {2006}}%
}]{%
Trotignon2006}
\APACinsertmetastar {%
Trotignon2006}%
\begin{APACrefauthors}%
Trotignon, J\BHBI G.%
, Mazelle, C\BPBI X.%
, Bertucci, C\BPBI L.%
\BCBL {}\ \BBA {} Acu{\~n}a, M\BPBI H.%
\end{APACrefauthors}%
\unskip\
\newblock
\APACrefYearMonthDay{2006}{}{}.
\newblock
{\BBOQ}\APACrefatitle {Martian shock and magnetic pile-up boundary positions
  and shapes determined from the Phobos~2 and Mars Global Surveyor data sets}
  {Martian shock and magnetic pile-up boundary positions and shapes determined
  from the phobos~2 and mars global surveyor data sets}.{\BBCQ}
\newblock
\APACjournalVolNumPages{Planet. Space Sci.}{54}{4}{357--369}.
\PrintBackRefs{\CurrentBib}

\bibitem [\protect \citeauthoryear {%
Zhang%
, Dong%
, Zhou%
\BCBL {}\ \protect \BOthers {.}}{%
Zhang%
\ \protect \BOthers {.}}{%
{\protect \APACyear {2025}}%
}]{%
Zhang2025a}
\APACinsertmetastar {%
Zhang2025a}%
\begin{APACrefauthors}%
Zhang, C.%
, Dong, C.%
, Zhou, H.%
\BCBL {}\ \BOthersPeriod {.}\end{APACrefauthors}%
\unskip\
\newblock
\APACrefYearMonthDay{2025}{}{}.
\newblock
{\BBOQ}\APACrefatitle {Anomalous transient enhancement of planetary ion escape
  at Mars} {Anomalous transient enhancement of planetary ion escape at
  mars}.{\BBCQ}
\newblock
\APACjournalVolNumPages{Nature Communications}{16}{}{}.
\newblock
\begin{APACrefDOI} \doi{10.1038/s41467-025-58351-y} \end{APACrefDOI}
\PrintBackRefs{\CurrentBib}

\bibitem [\protect \citeauthoryear {%
Zhang%
\ \protect \BOthers {.}}{%
Zhang%
\ \protect \BOthers {.}}{%
{\protect \APACyear {2022}}%
}]{%
Zhang2022}
\APACinsertmetastar {%
Zhang2022}%
\begin{APACrefauthors}%
Zhang, C.%
\BCBT {}\ \BOthersPeriod {.}
\end{APACrefauthors}%
\unskip\
\newblock
\APACrefYearMonthDay{2022}{}{}.
\newblock
{\BBOQ}\APACrefatitle {Three-Dimensional Configuration of Induced Magnetic
  Fields Around Mars} {Three-dimensional configuration of induced magnetic
  fields around mars}.{\BBCQ}
\newblock
\APACjournalVolNumPages{Journal of Geophysical Research: Planets}{127}{}{}.
\newblock
\begin{APACrefDOI} \doi{10.1029/2022JE007334} \end{APACrefDOI}
\PrintBackRefs{\CurrentBib}

\bibitem [\protect \citeauthoryear {%
Zhang%
\ \protect \BOthers {.}}{%
Zhang%
\ \protect \BOthers {.}}{%
{\protect \APACyear {2024}}%
}]{%
Zhang2024}
\APACinsertmetastar {%
Zhang2024}%
\begin{APACrefauthors}%
Zhang, C.%
\BCBT {}\ \BOthersPeriod {.}
\end{APACrefauthors}%
\unskip\
\newblock
\APACrefYearMonthDay{2024}{}{}.
\newblock
{\BBOQ}\APACrefatitle {Energetic Oxygen Ion Beams in the Martian Magnetotail}
  {Energetic oxygen ion beams in the martian magnetotail}.{\BBCQ}
\newblock
\APACjournalVolNumPages{Geophysical Research Letters}{}{}{}.
\PrintBackRefs{\CurrentBib}

\bibitem [\protect \citeauthoryear {%
Zhang%
\ \protect \BOthers {.}}{%
Zhang%
\ \protect \BOthers {.}}{%
{\protect \APACyear {2025}}%
}]{%
Zhang2025b}
\APACinsertmetastar {%
Zhang2025b}%
\begin{APACrefauthors}%
Zhang, C.%
\BCBT {}\ \BOthersPeriod {.}
\end{APACrefauthors}%
\unskip\
\newblock
\APACrefYearMonthDay{2025}{}{}.
\newblock
{\BBOQ}\APACrefatitle {Observational Characteristics of Electron Distributions
  in the Martian Induced Magnetotail} {Observational characteristics of
  electron distributions in the martian induced magnetotail}.{\BBCQ}
\newblock
\APACjournalVolNumPages{Geophysical Research Letters}{52}{}{}.
\newblock
\begin{APACrefDOI} \doi{10.1029/2024GL113030} \end{APACrefDOI}
\PrintBackRefs{\CurrentBib}

\end{thebibliography}
\end{document}